\documentclass[aps, prd, twocolumn,tightenlines, notitlepage, superscriptaddress, nofootinbib, preprintnumbers, floatfix,showkeys,11pt,altaffilletter]{revtex4-2}

\usepackage{upgreek}

\usepackage[official]{eurosym}
\usepackage[normalem]{ulem}
\usepackage{amstext}
\usepackage{graphicx}
\usepackage{url}
\usepackage{color}
\usepackage{ulem}
\usepackage[version=4]{mhchem}
\usepackage[utf8]{inputenc}
\usepackage{fontawesome}
\usepackage{yfonts}

\pdfoutput=1
\usepackage{textcomp}
\usepackage{comment}
\usepackage{yfonts}
\usepackage{epsfig,amsfonts,mathrsfs,graphicx,color,slashed,multirow}
\usepackage{latexsym,graphicx,slashed,color,enumerate,url,cancel,gensymb}
\usepackage{textcomp}

\usepackage[x11names]{xcolor}
\usepackage[colorlinks,pdfstartview=FitV,breaklinks=true]{hyperref}
\usepackage{booktabs}
\usepackage{adjustbox}
\usepackage{textgreek} 
\usepackage{caption, subcaption} 
\usepackage{float}

\usepackage{lmodern}
\usepackage{ae,aecompl}
\usepackage{appendix}


\definecolor{vdrgreen}{rgb}{0.0, 0.6, 0.0}

\definecolor{persiangreen}{rgb}{0.0, 0.65, 0.58}
\definecolor{mediumpersianblue}{rgb}{0.0, 0.4, 0.65}

\makeatletter
    \newcommand{\colorboxed}[3][white]{\fcolorbox{#2}{#1}{\m@th$\displaystyle#3$}}
\makeatother
\DeclareMathOperator{\diag}{diag}

\AtBeginDocument{\hypersetup{citecolor=mediumpersianblue,linkcolor=mediumpersianblue,urlcolor=mediumpersianblue}}
\usepackage{appendix}

\begin{document}


\title{{\LARGE Invisible neutrino decay at long-baseline neutrino oscillation experiments}}
\author{Christoph A. Ternes}
\email{christoph.ternes@lngs.infn.it}
\affiliation{Istituto Nazionale di Fisica Nucleare (INFN), Laboratori Nazionali del Gran Sasso, 67100 Assergi, L’Aquila (AQ), Italy
}
\author{Giulia Pagliaroli}
\email{giulia.pagliaroli@lngs.infn.it}
\affiliation{Istituto Nazionale di Fisica Nucleare (INFN), Laboratori Nazionali del Gran Sasso, 67100 Assergi, L’Aquila (AQ), Italy
}

\begin{abstract}
We perform an updated analysis of long-baseline accelerator data in the framework of neutrino oscillations in presence of invisible neutrino decay. We analyze data from T2K, NOvA and MINOS/MINOS+ and show that the combined analysis of all experiments improves the previous bound from long-baseline data by approximately one order of magnitude.
\end{abstract}
\maketitle

\section{Introduction}
\label{sec:intro}

The presence of three neutrino oscillations is, at present, observed in solar, atmospheric, reactor and accelerator experiments requiring the presence of nonzero neutrino masses and mixing angles ~\cite{deSalas:2020pgw,Capozzi:2021fjo,Esteban:2020cvm}.
A nonzero neutrino mass automatically opens the possibility that, besides oscillating, neutrinos can decay. This idea, firstly proposed as a mechanism to solve the solar neutrino problem~\cite{Bahcall:1972my,Pakvasa:1972gz}, is deeply investigated in the literature.
Standard Model neutrino decays, both through radiative and non-radiative processes, already have strong cosmological bounds thanks to the high precision measurement of the cosmic microwave background~\cite{Mirizzi:2007jd, Agashe:2013eba}.
On the other hand, processes involving beyond Standard Model (BSM) physics such as 
\begin{equation}
\nu_i \rightarrow \nu+X   
\label{eq:decay}
\end{equation}
are much less constrained. 
Here $\nu_i$ ($i=1,2,3$) is a neutrino mass eigenstate with mass $m_{i}$, while $\nu$ and $X$ are particles in the final state. In particular, $\nu$ can correspond to one or more neutrinos and $X$ represents one or more non-observable particles, typically identified as scalar or pseudoscalar fields. The BSM decay in Eq.~\eqref{eq:decay} can be classified as i) {\it visible decay}, in which at least one neutrino in the final state is active; ii) {\it invisible decay}, in which all $\nu$ and $X$ are non-observable particles. In this last case, the final state neutrino particles are {\it sterile} neutrinos $\nu_s$.

We are interested in discussing invisible neutrino decay. In this case, a beam of (relativistic) $\nu_{i}$-neutrinos with lifetime $\tau_i$, is depleted due to the invisible neutrino decay by the factor
\begin{equation} 
D_i(L,E)=\exp\left(-\alpha_i \times L/E\right)\,,
\label{dec}
\end{equation}
where  $E$ is the neutrino energy, $L$ is the distance between the source and the detector, and 
\begin{equation} 
\alpha_i=\frac{m_i}{\tau_i}\, 
\end{equation} 
is the decay parameter.
It is evident that, for a given ratio $L/E$, the neutrino decay is only sensitive to decay parameters with $\alpha_i\geq E/L$. 

Limits on $\alpha_i$ have been derived using different neutrino sources. Astrophysical neutrinos, travelling long distances, are most promising for testing neutrino decay. Indeed, for electron antineutrinos, the most favourable combination is provided by the supernova SN1987A ($L=50\,{\rm kpc}$, $E\sim 20 \,{\rm MeV}$): the observation of electron antineutrinos in Kamiokande-II~\cite{Kamiokande-II:1987idp} and IMB~\cite{Bionta:1987qt} and Baksan~\cite{Alekseev:1988gp} yields the lower limit $\alpha_1 \sim \alpha_2 \leq 10^{-5}$ eV/s \cite{Frieman:1987as, Ivanez-Ballesteros:2023lqa}. 
High-energy neutrinos observed by IceCube \cite{IceCube:2013low} are, in principle, more powerful. However we need to identify the source position or the production process to bound the $\alpha_i$ exploiting the observed flavor ratio \cite{Pagliaroli:2015rca}. The recent discovery of the first steady-state source of high-energy neutrinos, NGC 1068 \cite{IceCube:2022der}, allowed to investigate this possibility. However the present event rates are too low and the uncertainties on emission model too large to allow to set a bound on neutrino decay \cite{Valera:2023bud}. Neutrino decay has been also suggested as an option to alleviate the tension between cascade and track events in IceCube high energy data~\cite{Denton:2018aml}.

Atmospheric neutrinos or long-baseline neutrino data have been used so far to bound the $\nu_3$ lifetime.
As a leading example, the combined analysis of Super-Kamiokande data together with long-baseline data from K2K and MINOS provides a lower bound on the decay parameter $\tau_3/m_3\geq9.3\times 10^{-11}$~s/eV at $99\%$ confidence level (C.L.) \cite{Gonzalez-Garcia:2008mgl}. Whereas, the combined analysis of NOvA and T2K initial data sets showed a small preference for the decay scenario with a best-fit of $\tau_3/m_3=3.16\times 10^{-12}$ s/eV. However, at $3\sigma$ the combined constraint was found to be $\tau_3/m_3\geq 1.5\times 10^{-12}$ s/eV~\cite{Choubey:2018cfz}.   

In this paper, we improve the bound on the $\nu_3$ lifetime by combining larger data sets from the MINOS/MINOS+, NOvA and T2K experiments under the assumption of neutrino decay into invisible daughters. 

\section{The formalism}
\label{sec:formalism}

Let us assume that active neutrinos are subject to both standard mixing and invisible decays. Therefore, propagating neutrinos mix among flavor eigenstates in an oscillatory time-reversible manner and disappear due to time-irreversible decay. In this hypothesis,
the evolution process is described by the Hamiltonian given by
\begin{equation}
 H = \frac{1}{2E} \left[H_0 + H_m + H_D\right],
 \label{Ham_decay}
\end{equation}
where the first two terms correspond to the standard  vacuum and  matter terms, 
\begin{eqnarray}
 H_0 &=& U\diag(0,\Delta m_{21}^2,\Delta m_{31}^2) U^\dagger\,,\\
 H_m &=& 
\diag(V,0,0)\,,
\end{eqnarray}
with $V=2E\sqrt{2}G_F N_e$, where $E$ is the neutrino energy, $G_F$ the Fermi constant and $N_e$ the electron number density. The last term in Eq.\ (\ref{Ham_decay}) represents the neutrino decay part
\begin{equation}
 H_D = U\diag(-i\alpha_1,-i\alpha_2, -i\alpha_3)
U^\dagger .
\end{equation}
In this paper we are interested in the decay of $\nu_3$ only, hence we set $\alpha_1=\alpha_2=0$.
Neutrino decay implies that the sum of the neutrino oscillation probabilities might be different from one,
\begin{equation}
 P_{\alpha e} + P_{\alpha \mu} +P_{\alpha \tau} < 1,\quad \alpha=e,\mu,\tau.
\label{eq:osc_sum}
\end{equation}
This is a feature which will be used in the analysis of MINOS/MINOS+ data below, where we include the neutral current data set, which would be affected due to Eq.~\eqref{eq:osc_sum}. For discussion of the neutrino oscillation probability or formulas derived in this scenario, we refer the interested reader to Refs.~\cite{Lindner:2001fx,Abrahao:2015rba,Ghoshal:2020hyo,Chattopadhyay:2021eba,Chattopadhyay:2022ftv,Banerjee:2023sxj}.

\section{Data analysis}
\label{sec:data}

In this section we describe the data that we use in the analyses of this paper. In our analyses we keep the solar parameters fixed at the best fit values from Ref.~\cite{deSalas:2020pgw}. 
In addition, we have checked that reactor experiments are only sensitive to values of $\alpha_3$ larger than the ones considered here. Therefore in our analyses we include priors on $\theta_{13}$ and $\Delta m_{31}^2$ as obtained from the combined analysis of Daya Bay~\cite{DayaBay:2018yms} and RENO~\cite{jonghee_yoo_2020_4123573} data~\cite{deSalas:2020pgw}, which read

\begin{eqnarray}
    \sin^2\theta_{13} &=& 0.0221 \pm 0.0007\,,\nonumber\\
    \Delta m_{31}^{2,\text{NO}} &=& (2.59\pm0.06)\times10^{-3}~\text{eV}^2\,,\nonumber\\
    \Delta m_{31}^{2,\text{IO}} &=& -(2.54\pm0.06)\times10^{-3}~\text{eV}^2\,.
    \label{eq:priors}
\end{eqnarray}
The determination of $\sin^2\theta_{13}$ is the same for normal and inverted neutrino mass ordering (NO and IO, respectively). In the following subsections we briefly detail the analysis strategy for MINOS/MINOS+, T2K, and NOvA, the experiments under consideration in this paper.

\subsection{MINOS/MINOS+}
The accelerator-based neutrino oscillation experiment MINOS studied neutrinos produced at Fermilab at the NuMI beam facility and detected them at two detectors placed at 1.04~km and 735~km distance. 
First the neutrino beam peaked at an energy of 3~GeV, which later was increased to a peak energy of 7~GeV. These phases were called MINOS and MINOS+, respectively. In the analysis of this paper we consider the full data set of both phases, corresponding to an exposure of $10.56\times10^{20}$ protons-on-target (POT) in MINOS and $5.80\times10^{20}$ POT in MINOS+~\cite{Adamson:2017uda}. 
We use the publicly available MINOS/MINOS+ code used in the analysis for the search for light sterile neutrinos, see Ref.~\cite{Adamson:2017uda}. We modify this code to account for neutrino decay instead of active-sterile neutrino oscillations. The statistical analysis is performed by using the following $\chi^2$ definition as a function of the oscillation parameter set $\vec{p}$: 
\begin{widetext}
\begin{equation}
    \chi^2_{\mathrm{MINOS}}(\vec{p}) = \sum_{\text{Det.}}\sum_\text{channels}\sum_{i,j} (N_{\text{dat},i} - N_{\text{exp},i}(\vec{p}))~[V_{\text{cov.}}]^{-1}_{ij}~(N_{\text{dat},j} - N_{\text{exp},j}(\vec{p})) + \chi^2_{\text{pen.}}\,,
\end{equation}
\end{widetext}
where $N_{\text{dat},i}$ and $N_{\text{exp},i}$ are the observed and the predicted event rates, in the energy bin $i$. The covariance matrix $V_\text{cov.}$ accounts for the contributions from several sources of systematic uncertainties~\cite{Adamson:2017uda}. The last term,  
\begin{equation}
    \chi^2_{\text{pen.}} = \frac{(\Delta m_{31}^2 - \Delta m_{31}^{2,X})^2}{\sigma(\Delta m_{31}^{2,X})^2} + \frac{(\theta_{13}-\theta_{13}^{\text{BF}})^2}{\sigma(\theta_{13}^{\text{BF}})^2}\,,
\end{equation}
where $X=\text{NO, IO}$ for normal (inverted) ordering, contains the priors discussed above, see Eq.~\eqref{eq:priors}. The first two sums are taken over detectors (near and far) and over the different channels used in MINOS/MINOS+, which include disappearance and appearance channels using charged current events, and also the neutral current data which is sensitive to the effect described in Eq.~\eqref{eq:osc_sum}.

\subsection{T2K and NOvA}
We also consider data collected by the modern experiments T2K~\cite{T2K:2023smv} and NOvA~\cite{NOvA:2021nfi}. 
Both experiments observe events due to interactions of neutrinos and antineutrinos. In the case of T2K an exposure at Super-Kamiokande of 1.97$\times10^{21}$ POT in neutrino mode and 1.63$\times10^{21}$ POT in antineutrino mode has been reached up to now. 
NOvA, instead, has reached 1.36$\times10^{21}$~POT in neutrino mode~\cite{NOvA:2018gge} and 1.25$\times10^{21}$~POT in antineutrino mode.
The analysis for these experiments is performed using GLoBES~\cite{Huber:2004ka,Huber:2007ji}. We assume Gaussian smearing for the energy reconstruction in both experiments. In addition we use bin-to-bin efficiencies, which are adjusted to reproduce the best-fit spectra reported by the T2K and NOvA collaborations.
Also included are systematic uncertainties to account for uncertainties in the predicted signal and background spectra, see Refs.~\cite{Abe:2021gky,T2K:2023smv} and~\cite{NOvA:2021nfi}.
In the case of T2K and NOvA the $\chi^2$ function is given by
\begin{widetext}
\begin{equation}
\label{eq:T2KNOvAchi2}
 \chi^2_{\mathrm{T/N}}(\vec{p})=\min_{\vec{\delta}}
 \sum_\text{channels}2\sum_i \left[ N_{\text{exp},i}(\vec{p},\vec{\delta})- N_{\text{dat},i} +
 N_{\text{dat},i} \ln \left(\frac{N_{\text{dat},i}}{N_{\text{exp},i}(\vec{p},\vec{\delta})}\right)\right] 
 + \sum_i \left(\frac{\delta_i}{\sigma_i}\right)^2 + \chi^2_{\text{pen.}}\,,
\end{equation}
\end{widetext}
where again $N_{\text{dat},i}$ and $N_{\text{exp},i}$ are the data and prediction in bin $i$ and the first sum is taken over the different oscillation channels: $\nu_{\mu}\to\nu_{\mu}$, $\overline{\nu}_{\mu}\to\overline{\nu}_{\mu}$, $\nu_{\mu}\to\nu_e$ and $\overline{\nu}_{\mu}\to\overline{\nu}_e$. The second term contains the penalties $\sigma_i$ for the systematic uncertainties $\delta_i$, while the last term is equivalent to the one in the $\chi^2$ function for MINOS/MINOS+. 

In addition to the individual analyses of the T2K, NOvA and MINOS/MINOS+ experiments, we will also discuss a combined analysis of all accelerator data. Note that in this case the penalty term $\chi^2_\text{pen.}$ is added only once to the overall $\chi^2$ function.

\section{Results and discussion}
\label{sec:results}

\begin{figure}
\centering
    \includegraphics[width=0.48\textwidth]{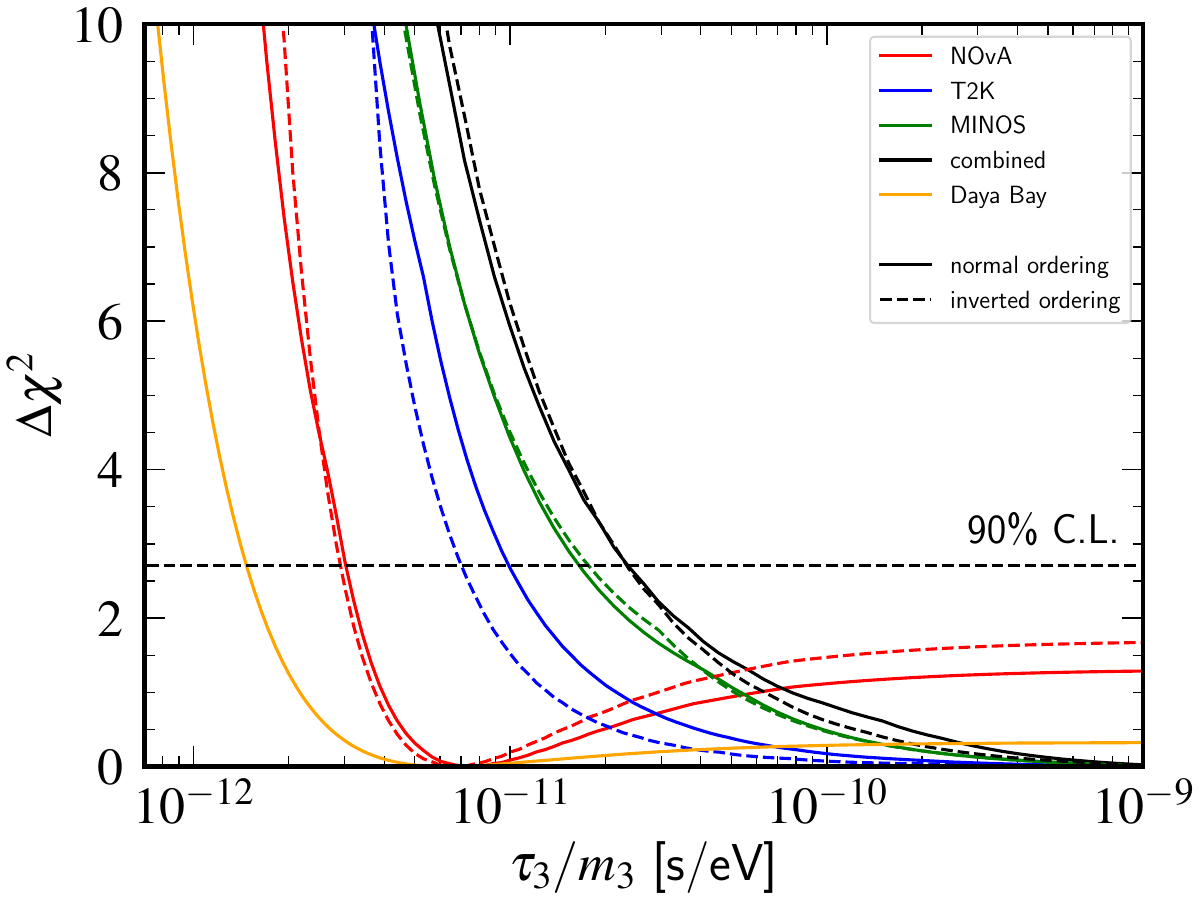}
\caption{The bounds on $\tau_3/m_3$ obtained from the analysis of NOvA (red), T2K (blue), MINOS/MINOS+ (green) data, and the result from the combined fit (black). The solid (dashed) lines correspond to the analysis using normal (inverted) neutrino mass ordering. Also shown is the bound from Daya Bay (orange), which, as explained in the text, is weaker.}
\label{fig:lifetime_bound}
\end{figure}

In this section we discuss the results obtained in this paper. When performing the analyses we always consider either normal or inverted neutrino mass ordering. When plotting the results they are always plotted with respect to the best fit value obtained in each ordering separately. The main result is presented in Fig.~\ref{fig:lifetime_bound}. Apart from long-baseline bounds we also show the bound from an analysis of Daya Bay data. The analysis procedure is the same as in Ref.~\cite{DeRomeri:2023dht}, but with the decoherence probability replaced with the one including neutrino decay. As anticipated in Sec.~\ref{sec:data}, reactor experiments are sensitive to smaller (larger) values of $\tau_3/m_3$ ($\alpha_3$). This justifies to include the priors of Eq.~\eqref{eq:priors} in our analyses of accelerator data. The different colors correspond to the analyses of the individual experiments, while solid (dashed) lines are obtained using normal (inverted) neutrino mass ordering. As can be seen, the $\chi^2$ profiles are very similar for both orderings. The only major difference appears in the analysis of T2K data, where the bound for IO is weaker than the bound for NO. This happens, because in the case of inverted neutrino mass ordering T2K data prefer quite small values of $|\Delta m_{31}^2|$, see Ref.~\cite{Abe:2021gky}. Therefore, there is a tension between T2K data (in inverted ordering) and the reactor prior imposed in our analysis. This does not happen in normal ordering, where there is better agreement. Therefore the analysis in IO is penalized resulting in a weaker bound.
Note that in the case of NOvA we find a best fit value with finite lifetime. As can be seen from the profile, this best fit is statistically not very significant, and it disappears after combining with the other experiments.

\begin{figure}
\centering
    \includegraphics[width=0.48\textwidth]{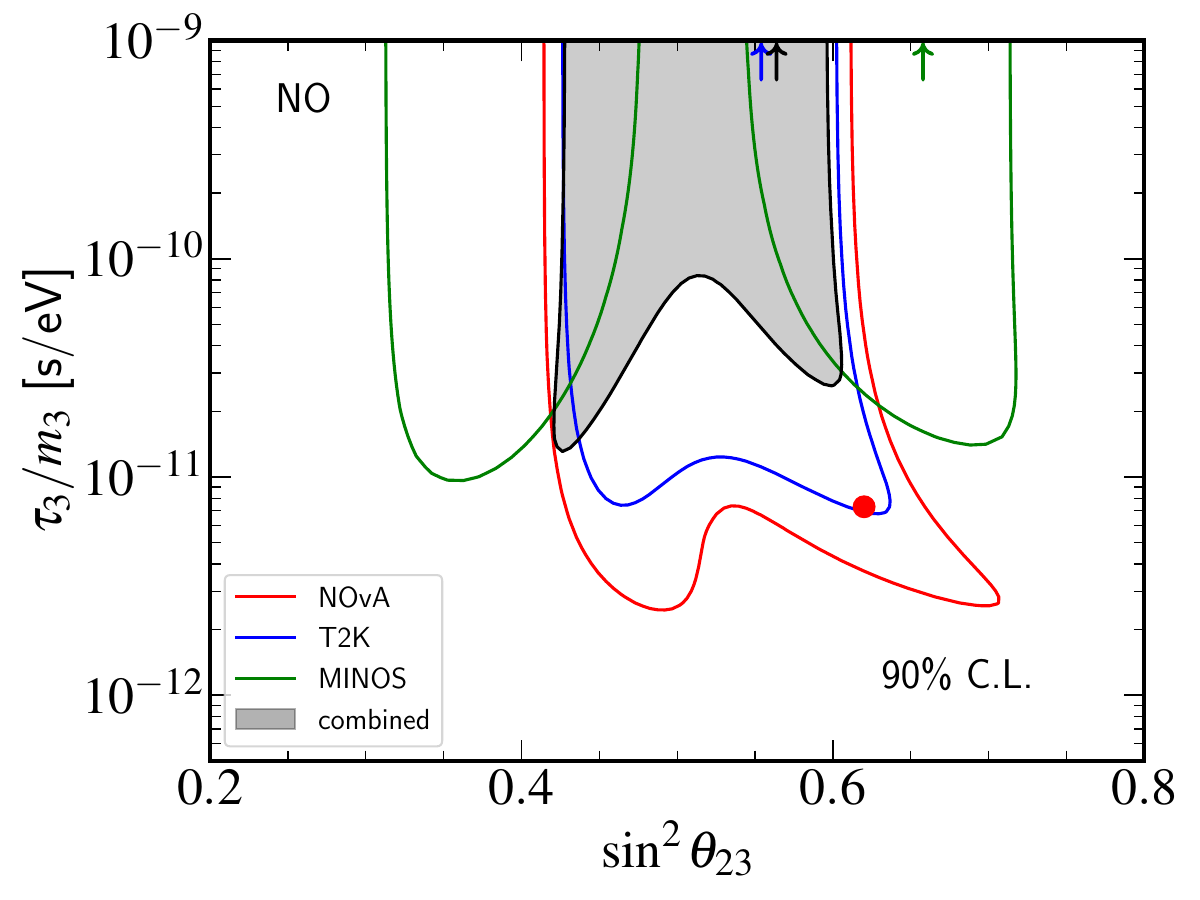}
    \includegraphics[width=0.48\textwidth]{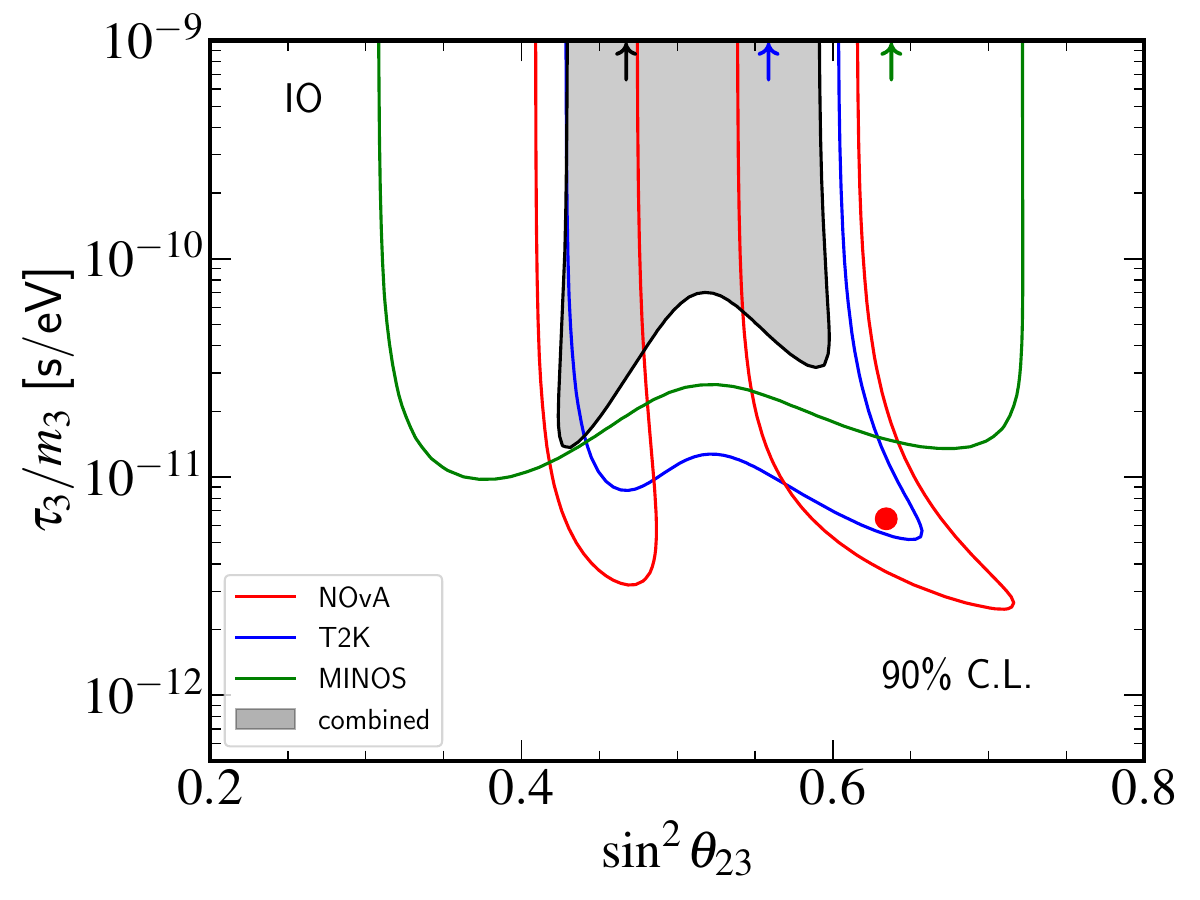}
\caption{Contours in the $\tau_3/m_3-\sin^2\theta_{23}$-plane at 90\% confidence level for two degrees of freedom for normal (upper panel) and inverted (lower panel) neutrino mass ordering. We show the contours from the analysis of T2K (blue), NOvA (red), MINOS/MINOS+ (green) data and from the combination (shaded region). The arrows and the dot indicate the best fit values found in each analysis.}
\label{fig:sq23_lifetime}
\end{figure}

In Fig.~\ref{fig:sq23_lifetime} we show the contours in the $\tau_3/m_3-\sin^2\theta_{23}$-plane at 90\% confidence level for two degrees of freedom for the different experiments (colored lines) and for the combined analysis (shaded region). From this figure we see that there is a correlation between $\tau_3/m_3$ and $\sin^2\theta_{23}$. 
Note, in particular in the contours for MINOS/MINOS+, that values of $\sin^2\theta_{23}$ away from maximal mixing result in weaker bounds on $\tau_3/m_3$. This is the reason why the combined analysis of all data provides a better bound than MINOS/MINOS+ data alone. The inclusion of NOvA and T2K data pushes the allowed region in $\sin^2\theta_{23}$ closer to maximal mixing\footnote{At this point it should be noted that the results of this paper are not in tension with the ones in Ref.~\cite{deSalas:2020pgw}. The reason why in the present analysis MINOS/MINOS+ data disfavor maximal mixing, while it is allowed in Ref.~\cite{deSalas:2020pgw}, is the inclusion of the prior in Eq.~\eqref{eq:priors}. In Refs.~\cite{deSalas:2020pgw,Adamson:2017uda}, $\Delta m_{31}^2$ is also fitted with the data resulting in a different behavior. The same argument applies to the analyses of T2K and NOvA data. We always make sure that our analyses reproduce the official results when making the same assumptions as in the original references.} which results in a stronger bound on $\tau_3/m_3$. This feature can also be seen by observing how the best fit point from MINOS/MINOS+, indicated by the green arrow in Fig.~\ref{fig:sq23_lifetime}, is outside of the allowed parameter space from T2K and NOvA. Therefore, upon combining all data we are left with a more restrictive parameter space for $\tau_3/m_3$.

We summarize the bounds that we obtain in this paper in Tab.~\ref{tab:bounds}. Our analyses of NOvA and T2K data update the analysis performed in Ref.~\cite{Choubey:2018cfz}, while the MINOS analysis updates Ref.~\cite{Gomes:2014yua}. The individual bounds from NOvA, T2K and MINOS/MINOS+ data improve the previous bounds all by a factor of 3--4. 
From our combined analysis we find that long-baseline accelerator data provide a bound of

\begin{equation}    
    \tau_3/m_3 \geq 2.4\times 10^{-11}~\text{s/eV at 90$\%$ C.L.}
    \label{eq:bound}
\end{equation}
for both neutrino mass orderings. 
Therefore, we find that our global analysis of long-baseline accelerator data improves the previous bound from Ref.~\cite{Choubey:2018cfz} ($\tau_3/m_3 \geq 2.4\times 10^{-12}$~s/eV at 90$\%$ C.L.) by one order of magnitude. Note that, unlike in our case, no prior on $\Delta m_{31}^2$ was used in Ref.~\cite{Choubey:2018cfz}. However, we have verified that the bound obtained in this analysis, Eq.~\eqref{eq:bound}, does not depend on this prior. The same bound is obtained when allowing $\Delta m_{31}^2$ to vary freely in the analysis. This is expected since the determination of $\Delta m_{31}^2$ from long-baseline accelerator data is more precise than the one from reactor data.

As a final remark, let us mention that invisible neutrino decay does not have any impact on the preferred neutrino mass ordering. The preference obtained from the combined analysis is the same as the one obtained in the standard analysis. In addition, we also found that invisible neutrino decay neither helps to alleviate the disagreement in the measurement of $\delta$ in T2K and NOvA.

\begin{table}[t]
    \centering
    \begin{tabular}{|c|c|c|}
\hline
         Experiment & NO & IO \\\hline
         ~~NOvA~~ & $3.0\times10^{-12}$~s/eV & $2.9\times10^{-12}$~s/eV \\
         ~~T2K~~ &           $9.9\times10^{-12}$~s/eV &          $7.0\times10^{-12}$~s/eV \\
         ~MINOS~ & $1.6\times10^{-11}$~s/eV & $1.8\times10^{-11}$~s/eV \\
         combined &  $2.4\times10^{-11}$~s/eV & $2.4\times10^{-11}$~s/eV \\
         \hline
    \end{tabular}
    \caption{\label{tab:bounds} The bounds at 90\% C.L. on $\tau_3/m_3$ obtained in the analyses of this paper.}
\end{table}
\section{Conclusions}
\label{sec:conc}

We have performed an analysis of long-baseline accelerator data to place a new bound on invisible neutrino decay. We find that the bound is driven by MINOS/MINOS+ data, and then further improved by the inclusion of NOvA and T2K data. This improvement happens due to the partial breaking of a correlation between $\sin^2\theta_{23}$ and $\tau_3/m_3$ in the MINOS/MINOS+ analysis. Basically, the inclusion of NOvA and T2K data drives $\sin^2\theta_{23}$ closer towards maximal mixing, which improves the bound on $\tau_3/m_3$. 

The individual up-to-date bounds on $\tau_3/m_3$ obtained in this analysis improve the former bounds from Refs.~\cite{Gomes:2014yua,Choubey:2018cfz} by a factor of 3--4. A stronger bound than the one obtained here has been reported in Ref.~\cite{Gonzalez-Garcia:2008mgl} using an older MINOS dataset and also data from Super-Kamiokande. However, it should be noted that the bound in this reference shows non-Gaussian behavior due to the presence of a local minimum. Not taking into account this bound, Eq.~\eqref{eq:bound} is the strongest bound on the neutrino lifetime obtained from neutrino oscillation data so far\footnote{Note that very strong bounds on $\tau_3/m_3$ have been derived in Refs.~\cite{Funcke:2019grs,Picoreti:2021yct} for a specific decay model using solar and KamLAND data.}. 

This laboratory bound will be further improved at future facilities. It has been shown in Refs.~\cite{Choubey:2017dyu,Ghoshal:2020hyo,Chakraborty:2020cfu} that DUNE and T2HK can improve our bound by a factor of about 2. A bound of similar strength is also expected at the JUNO experiment~\cite{Abrahao:2015rba,JUNO:2021ydg}. The future facility ESS$\nu$SB would only produce a bound of similar strength as the current one~\cite{Choubey:2020dhw}. Even stronger bounds can be expected from future atmospheric experiments, as shown in Refs.~\cite{Choubey:2017eyg,deSalas:2018kri,KM3NeT:2023ncz}.

Finally, we must remark that terrestrial experiments are not the only way to obtain bounds on invisible neutrino decay. Constraints on the neutrino lifetime have also been derived from astrophysical and cosmological observations, see e.g. Refs.~\cite{Beacom:2004yd,Fogli:2004gy,Hannestad:2005ex,Archidiacono:2013dua,Escudero:2019gfk,Barenboim:2020vrr,Pompa:2023yzg}. These are signiﬁcantly more stringent, up to a level~\cite{Barenboim:2020vrr} of $\tau/m_\nu > 8\times10^6$~s/eV, than the bounds obtained here or that can be expected from any other current or future neutrino oscillation experiment.
\section*{Acknowledgements}
The work of GP is partially supported by  grant number 2022E2J4RK "PANTHEON: Perspectives in Astroparticle and
Neutrino THEory with Old and New messengers" under the program PRIN 2022 funded by the Italian Ministero dell’Universit\`a e della Ricerca (MUR) and by the European Union – Next Generation EU.


%

\end{document}